\documentclass[reprint,amsmath,amssymb,aps,prl]{revtex4-1}

\usepackage{hyperref}
\usepackage{color}
\hypersetup{
    colorlinks=true,
    linkcolor=blue,
    citecolor=blue
}

\usepackage{graphicx}
\usepackage{dcolumn}
\usepackage{bm}
\usepackage[cp1251]{inputenc}

\begin{document}

\title{Observation of nonclassical correlations in biphotons generated from \\ an ensemble of pure two-level atoms}

\author{Michelle O. Ara\'ujo, Lucas S. Marinho, and Daniel Felinto}
 \affiliation{Departamento de F\'{\i}sica, Universidade Federal de Pernambuco, 50670-901, Recife, Pernambuco, Brazil}

\date{\today}

\begin{abstract}
We report the experimental verification of nonclassical correlations for an unfiltered spontaneous four-wave-mixing process in an ensemble of cold two-level atoms, confirming theoretical predictions by Du {\it et al.} in 2007 for the violation of a Cauchy-Schwarz inequality in the system, and obtaining $R = (1.98\pm0.03) \nleq 1$. Quantum correlations are observed in a nano-seconds timescale, in the interference between the central exciting frequency and sidebands dislocated by the detuning to the atomic resonance. They prevail over the noise background coming from Rayleigh scattering from the same optical transition. These correlations are fragile with respect to processes that disturb the phase of the atomic excitation, but are robust to variations in number of atoms and to increasing light intensities.
\end{abstract}

\pacs{Valid PACS appear here}

\maketitle

Four-wave mixing (FWM) is a nonlinear parametric process in which a fourth optical field is generated in a medium as a result of the action of three other fields~\cite{Yariv1989}. Commonly, two pump beams are employed to enhance a weak probe beam simultaneously with the generation of the FWM signal. In atomic systems, FWM has been an important source of squeezing and non-classical correlations for light fields (probe and signal) for more than three decades~\cite{Slusher1985,Maeda1987,Raizen1987,Vallet1990,Lambrecht1996,Ries2003,McCormick2007}. Since 2001, with the introduction of the Duan-Lukin-Cirac-Zoller (DLCZ) protocol for long distance quantum communication~\cite{Duan2001}, quantum correlations from spontaneous four-wave mixing (SFWM) have been recognised as a critical resource for quantum information applications. In SFWM, spontaneous emission in a particular direction plays the role of the weak probe field. Several groups in the last fifteen years have demonstrated the capability of SFWM to generate highly correlated photon pairs~\cite{Kuzmich2003,Balic2005,Matsukevich2005,Laurat2006,Thompson2006,Zhao2009, Albrecht2015,OrtizGutierrez2018}, and even applied it to implement crucial portions of the DLCZ protocol~\cite{Chou2005,Chou2007}. 

In order to observe quantum correlations in SFWM, various techniques have been employed to filter the correlated photons from the large backgrounds coming from elastic Rayleigh scattering of the pump beams. Rayleigh scattering is a ubiquitous process in which small particles scatter light with frequencies close to that of the excitation field~\cite{Strutt1871}. For any particular transition between two levels in an ensemble of atoms, this process results in scattered light with photon statistics of a thermal light source~\cite{Loudon1983,Moreira2021}. To eliminate this contribution, SFWM experiments typically employ a more complex level structure for the atoms, generating light with polarisations or frequencies significantly different to the pump beams.

In 2007, however, Du {\it et al.} pointed out theoretically that quantum correlations in SFWM from an ensemble of two-level atoms should overcome the Rayleigh scattering background, revealing its non-classical nature even without the use of any filter~\cite{Du2007,Wen2007,Wen2008}. This result was surprising for a number of reasons. First, a non-classical signal was surpassing a central process for classical optics under relatively mundane conditions. Second, it is generally assumed that quantum correlations in SFWM in free space have a better chance of prevailing at high number of atoms and low pump intensities, since these factors would enhance, respectively, the parametric process behind FWM~\cite{Felinto2005,deOliveira2014} and the purity of the quantum states~\cite{Duan2001,Laurat2006}. However, the non-classical correlations predicted by Du {\it et al.} were largely insensitive to both pump intensity and number of atoms in the ensemble, indicating its robustness for a large range of parameters. Third, the theory in Refs.~\onlinecite{Du2007,Wen2007} was deduced in a limit particularly suitable to classical models~\cite{Allen1987}, serving as first approximation for a large number of systems, which highlights the necessity of quantum theory to explain the whole body of phenomena for any range of parameters.

Unfortunately, the overall message of the works by Du {\it et al.} was significantly undermined by the lack of experimental demonstration of these non-classical correlations on the accompanying experiments~\cite{Du2007,Wen2007}. In the present work, we finally close this gap and report a series of experiments confirming the core claims of the theory in Refs.~\onlinecite{Du2007,Wen2007}. We observe not only the non-classical correlations without any filter for the Rayleigh scattering background, but also its insensitivity with variations in pump power and number of atoms. This opens numerous possibilities. First, if the correlations were strong enough to be observed without filters, we expect to enhance it significantly, in the future, by filtering the light component at the pump frequency. Second, multipartite quantum correlations should play an important role in this system, since its symmetry allows for SFWM in two other directions~\cite{Lopez2019,Capella2021} not explored here or in the previous theoretical works. Finally, the independence of the effect with pump power and number of atoms may favour future applications, since it allows for more amenable conditions of moderate optical depths and higher rates of biphoton generation.

\begin{figure}[th]
    \centering
    \includegraphics[scale=0.40]{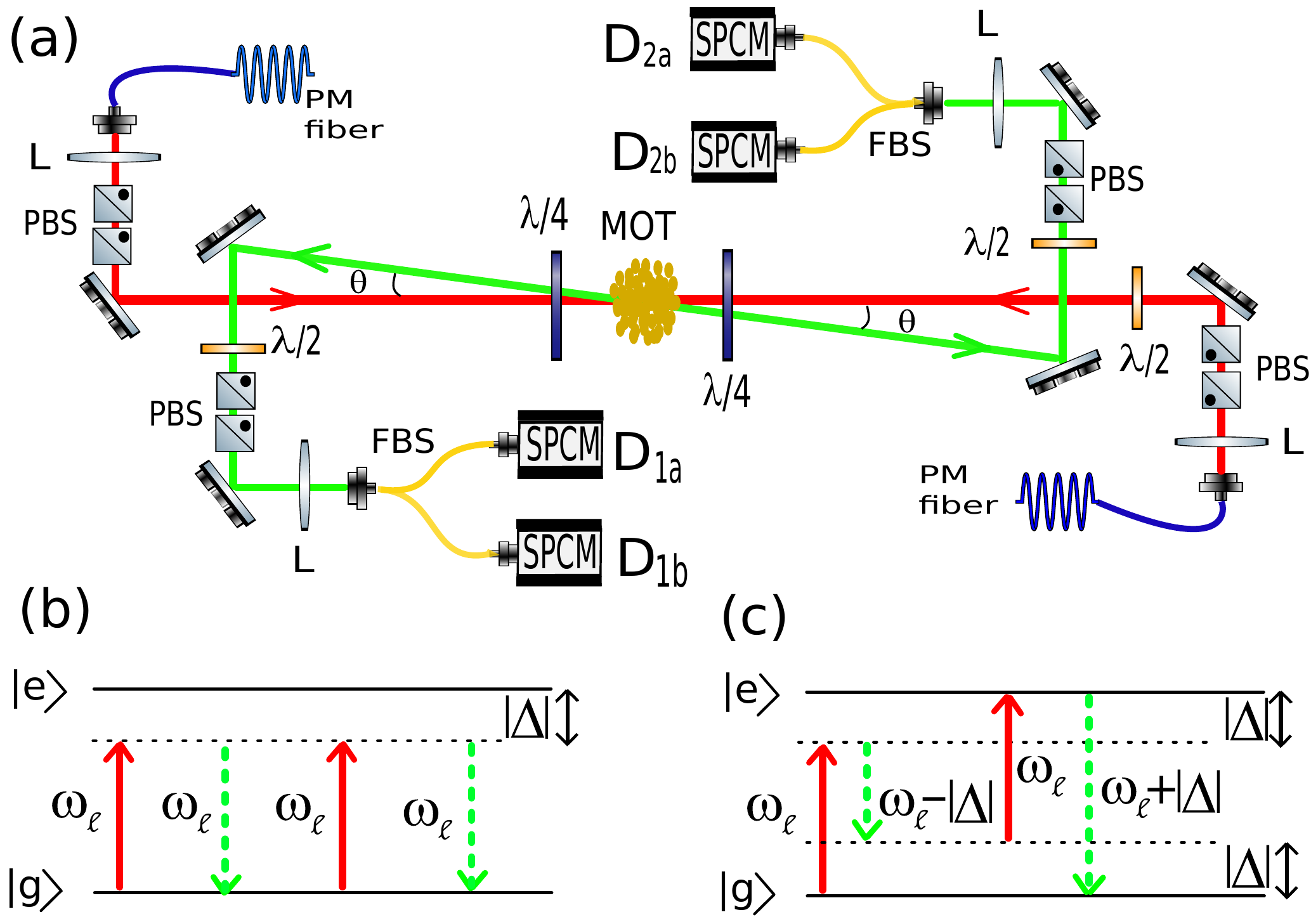}
        \caption{(a) Experimental setup for spontaneous four-wave mixing (SFWM) from a cold ensemble of pure two-level atoms. (b) and (c) show the two possible pathways for SFWM. MOT: magneto-optical trap; PBS: polarising beam splitter; SPCM: single-photon counting model; FBS: fiber beam splitter; L: lens; $\lambda/2$ ($\lambda/4$): half-wave (quarter-wave) plate; PM fiber: polarisation-maintaining fiber.}
    \label{Fig1}
\end{figure}

The experimental scheme under consideration is shown in Fig.~\ref{Fig1}a, with two counter propagating pump fields (in red) spontaneously generating pairs of photons (fields 1 and 2, in green) emitted in opposite directions, forming an angle $\theta$ with the pump fields. This backward SFWM process occurs mainly through two different pathways: either the emitted photons have frequencies $\omega_1 = \omega_2 = \omega_l$ (Fig.~\ref{Fig1}b), with $\omega_l$ the excitation-laser frequency, or they have frequencies $\omega_1 , \omega_2 = \omega_l \pm \Delta$ (Fig.~\ref{Fig1}c), with $\Delta = \omega_l - \omega_a$ and $\omega_a$ the atomic resonance frequency. These different contributions interfere in the observation of fields 1 and 2, resulting in a beating in the intensity cross-correlation function $g_{12}(t,t+\tau) = \langle I_{D1}(t) I_{D2}(t+\tau)\rangle / \langle I_{D1}(t)\rangle \langle I_{D2}(t+\tau)\rangle$, with $I_{Di}(t)$ the intensity of light at time $t$ measured at a detector for field $i$ ($i=1,2$), and $\langle \cdots \rangle$ denoting an ensemble average over many samples. Assuming an ergodic system and in the simple limit of large detunings, short delays, and low excitation power, the theoretical analysis of Refs.~\onlinecite{Du2007} and~\onlinecite{Wen2007} leads to
\begin{equation}
g_{12}(\tau) = 1 + \frac{4}{\pi^2} \left[ 1 + {\rm e}^{-\Gamma |\tau|} - 2 \cos \left( \Delta |\tau|\right) {\rm e}^{-\Gamma |\tau|/2} \right] \,, \label{g12}
\end{equation}
with $\Gamma$ the natural decay rate of the excited state. The maximum value of this quantity is $g_{12}^{max} \approx 2.62$. The fact that $g_{12}^{max} > 2$ for the cross-correlation function is quite significant for photon pair generation by SFWM. The corresponding auto-correlation functions are expected to be $g_{11} , g_{22} \approx 2$ due to the field's thermal statistics. The simplest measure of quantum correlation for biphoton generation in SFWM is the observation of $g_{12} > g_{11},g_{22}$~\cite{Kuzmich2003,Balic2005,Matsukevich2005,Laurat2006,Thompson2006,Zhao2009,OrtizGutierrez2018}, since this condition results in the violation of a Cauchy-Schwarz inequality valid for all classical fields~\cite{Clauser1974}. This observation essentially contradicts any theory for which the probability of photo-detection is proportional to the classical intensity of light. 

As anticipated above, Eq.~\eqref{g12} does not depend on pump power or the number of atoms, since the product of independent Rayleigh scatterings in fields 1 and 2 depends on these quantities in a similar way as the FWM process itself~\cite{Wen2007}, rendering normalised quantities such as $g_{12}$ insensitive to these parameters. The Rayleigh scattering is an elastic process, with the scattered photon having the same energy as a pump photon. However, it is not phase matched, since the change in direction of the photon implies in a change of linear momentum of the small scatterer. SFWM, on the other hand, is a phase-matching process, with both momentum and energy conservation among the light fields. In the situation considered here, thus, the phase matching condition enhances the efficiency of the process enough to surpass the product of two independent Rayleigh scatters. 

Our ensemble is a cold cloud of Rb$^{87}$ atoms obtained from a magneto-optical trap. The trap laser and magnetic field are turned off 1$\,$ms prior to turning on the pump fields, but the repumper laser is kept on for an extra 900$\,\mu$s to allow for the preparation of all atoms in the 5S$_{1/2}$($F$=2) hyperfine ground state~\cite{Moreira2021}. 50$\,\mu$s after turning off the repumper laser, counter-propagating pump fields of same intensity $I_p$ are turned on for 1$\,$ms. After that, the trap is turned on again and the cycle is repeated at a 40$\,$Hz rate. As shown in Fig.~\ref{Fig1}a, the pump fields have their linear polarisations transformed to the same circular $\sigma^+$ polarisation by quarter-wave ($\lambda/4$) plates on each side of the atomic ensemble, resulting in optical pumping to the 5S$_{1/2}$($F=2,m_F$=2) Zeeman sub-level~\cite{deAlmeida2016,SM}. We collect fields 1 and 2 using single mode fiber, with $\theta = 3^{\circ}$. Prior to the fibers, the $\lambda/4$ plates transform the circular polarisation of the emitted photons back to linear polarisation. The photon's degree of polarisation is verified to be $(99.0\pm 0.2)$\%. The 4$\sigma$ beam diameters at the ensemble are $420\,\mu$m and $140\,\mu$m for pump fields and detection modes, respectively. The optical depth is $OD \approx 15$, corresponding to $N \approx 10^6$ atoms contributing to light in the detected modes~\cite{OrtizGutierrez2018,deOliveira2014}. $\Delta$ is the detuning from the excited state 5P$_{3/2}$($F=3$). The two emitted fields go through fiber beam splitters towards two separate avalanche photodetectors for fields 1 ($D_{1a}, D_{1b}$) and 2 ($D_{2a},D_{2b}$), respectively. The detectors are single-photon counting modules (model SPCM-AQRH-13-FC from Perkin Elmer) with outputs directed to a multiple-event time digitizer with 100$\,$ps time resolution (model MCS6A from FAST ComTec). 

The relatively small nonclassical correlations coming from Eq.~\eqref{g12}, when compared to the state of the art of SFWM~\cite{Laurat2006}, imply that these are more fragile, easier to lose. For this reason, the first attempts to measure these nonclassical correlations were not successful~\cite{Du2007,Wen2007}. The experimental observation of non-classical correlations reported here was possible due to various improvements with respect to the apparatus and methods of Refs.~\onlinecite{Du2007,Wen2007}. The first was to optically pump the atoms towards a single Zeeman sub-level, resulting in a pure two-level system, since atoms in different Zeeman sub-levels participate in different parametric processes and do not constructively interfere with each other to enhance the SFWM~\cite{Felinto2005}. The second improvement was to turn off the magnetic field and repumper laser during trials, since they could scramble the collective phase throughout the atomic ensemble. Particularly, turning off the repumper laser implies that our system is no longer ergodic, since we observe now a clear temporal dynamics related to optical pumping. Thus, the methodology to measure the effect behind Eq.~\eqref{g12} had to change, focusing in ensemble averages rather than time averages. Finally, the time resolution of the coincidence electronics was improved by a factor of 10, allowing us to probe faster oscillations for larger detunings.

\begin{figure}[h]
    \centering
    	\includegraphics[scale=0.4]{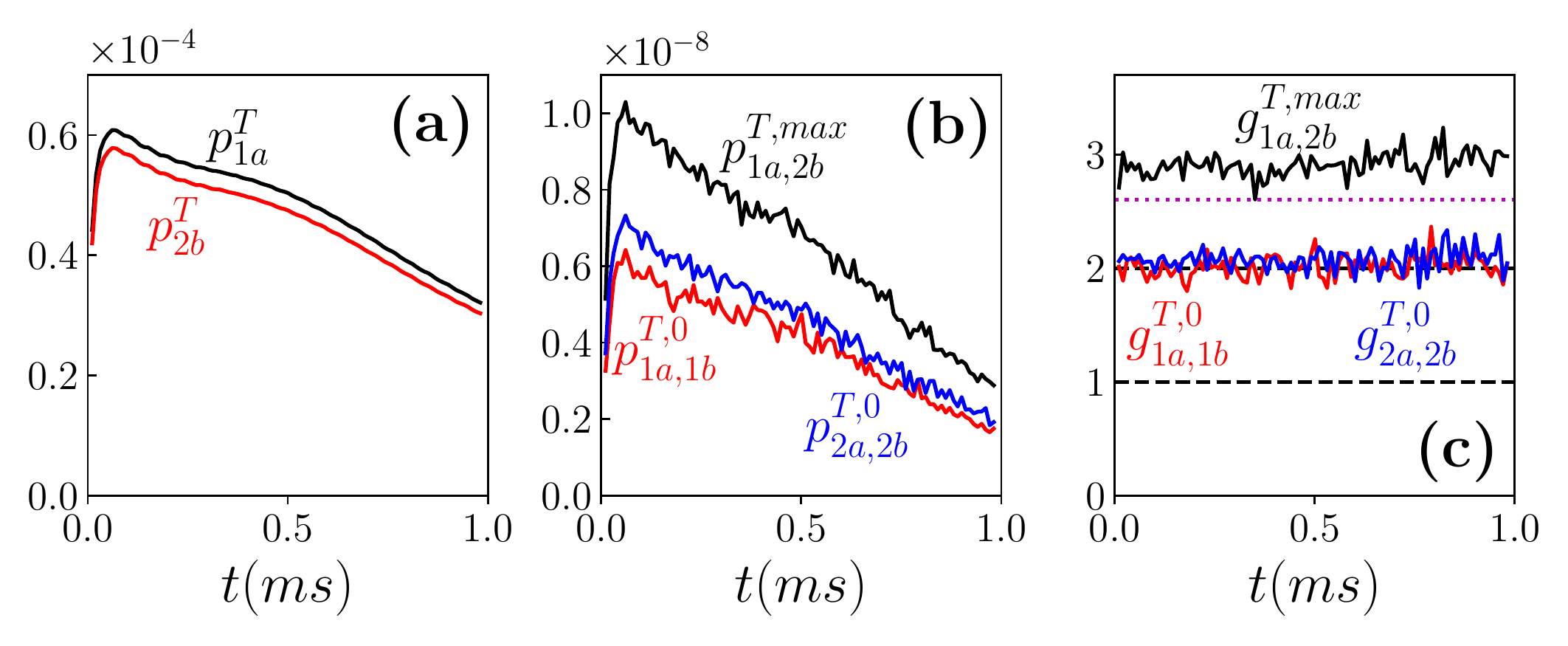}
	\vspace{-0.6cm}
        \caption{(a) Probabilities for single detections and (b) maximum joint detections as a function of time, with averages over $T = 10\,\mu$s. (c) Normalized second order correlation functions for the joint probabilities in (b). The dashed lines indicate the levels of 1 for no correlation, 2 for thermal correlations, and 2.62, for the maximum of Eq.~\eqref{g12}.}
    \label{Fig2}
\end{figure}

As just mentioned, during each 1$\,$ms trial period, the detection probabilities vary due to spurious optical pumping to the 5S$_{1/2}$($F$=1) level, coming from the action of a residual magnetic field of about 23~mG that weakly couples the atoms to non-cycling transitions~\cite{Moreira2021,SM}. This is shown in Fig.~\ref{Fig2}a, where we plot two of the four single detection probabilities, namely, $p_{1a}^T(t)$ and $p_{2b}^T(t)$ for detectors $D_{1a}$ and $D_{2b}$, respectively. These probabilities were first obtained as ensemble averages for each time $t$ over all trial periods, and later averaged over a time interval of $T=10\,\mu$s around $t$.  Considering that Rayleigh scattering would result in $p_{1a}^T,p_{2b}^T \propto N$, we observe then an initial period of optical pump in the Zeeman structure, as atoms accumulate in the $m_F = +2$ sub-level. Later, the spurious optical pump to the hyperfine $F=1$ level slowly decreases the overall population in $F=2$. For Fig.~\ref{Fig2}, $I_p = 126\,$mW/cm$^2$ (measured power per area) and $\Delta = +20\,\Gamma$. 

\vspace{-0.2cm}

Figure~\ref{Fig2}b plots three of the joint probabilities $p_{ij}^T$, between detectors $i$ and $j$, characterising the biphoton correlations. For each of them, we plotted their maxima as $\tau$ is varied for a particular instant $t$, given by $p_{i,j}^{T,max}$. In the case of joint detections for the same field, their maxima occur at $\tau = 0$, represented by $p_{1a,1b}^{T,0}(t)$ and $p_{2a,2b}^{T,0}(t)$, respectively. As expected, joint probabilities are more sensitive to the number of atoms than single probabilities, since even uncorrelated detections would give $p_i^Tp_j^T \propto N^2$. Most importantly, however, is that $p_{1a,2b}^{T,max}(t)$ is always larger than $p_{1a,1b}^{T,0}(t)$ and $p_{2a,2b}^{T,0}(t)$, indicating already the presence of non-classical correlations. As previously discussed, these correlations are better quantified through the corresponding normalised functions $g_{ij}^{T,max}(t)$ of Fig.~\ref{Fig2}c. We observe then $g_{1a,1b}^{T,0}(t) \approx g_{2a,2b}^{T,0}(t) \approx 2$, as expected for fields with thermal statistics~\cite{Moreira2021,Eloy2018}. On the other hand, we have always $g_{1a,2b}^{T,max}(t) > 2$.

As predicted from Eq.~\eqref{g12}, $g_{1a,2b}^{T,max}(t)$ is almost independent of the number of atoms. We observe this trend for all correlation functions and for every $\tau$. In this way, we can capture well the behaviour of the system by averaging the correlation functions during the whole 1ms trial period, defining $\bar{g}_{ij}(\tau) = g_{ij}^{T=1\,{\rm ms}}(0.5\,{\rm ms},0.5\,{\rm ms}+\tau)$. The results of Fig.~\ref{Fig2} were acquired with enough statistics to determine any correlation function or the violation of the Cauchy-Schwarz inequality throughout the whole 1$\,$ms trial period. On the other hand, the averaged $\bar{g}_{ij}(\tau)$ allows for a broader variation of parameters in the system, without the need to accumulate as much data. We can then define the corresponding averaged Cauchy-Schwarz inequality for classical fields as~\cite{Clauser1974}
\begin{equation}
R(\tau) = \frac{\bar{g}_{1a,2b}(\tau)\bar{g}_{1b,2a}(\tau)}{\bar{g}_{1a,1b}(0)\bar{g}_{2a,2b}(0)} \leq 1 \,. \label{R}
\end{equation}

\begin{figure}[h]
    \centering
        \includegraphics[scale=0.42]{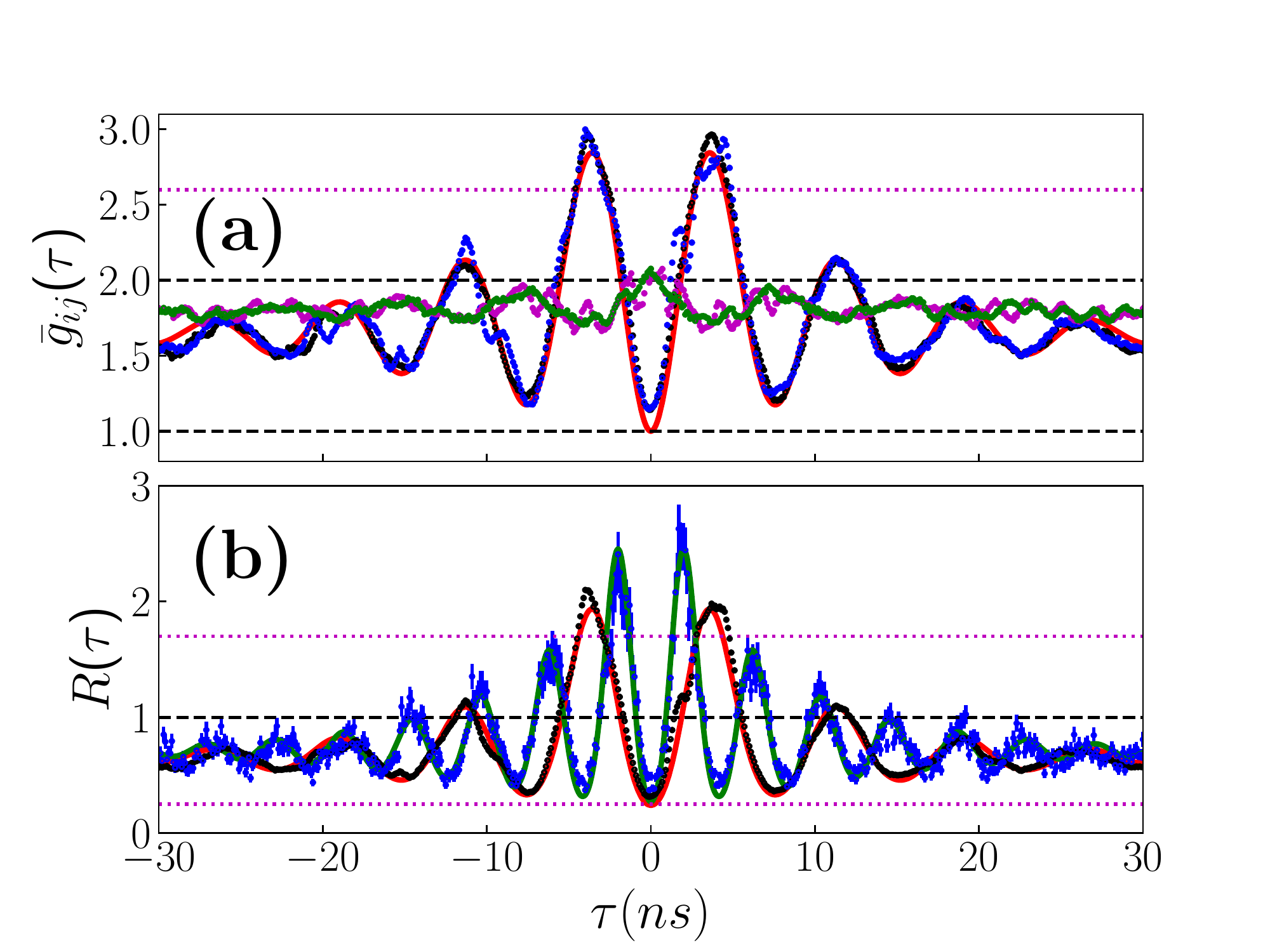}
        \vspace{-0.4cm}
        \caption{(a) Circles provide measurements of the second order correlation functions, as a function of $\tau$, between detectors $1a$ and $2b$ (black), $1b$ and $2a$ (blue), $1a$ and $1b$ (magenta), $2a$ and $2b$ (green). Parameters and dashed lines are the same as for Fig.~\ref{Fig2}. Red solid line provides a fit to Eq.~\eqref{g12b}. (b) Circles provide measurements of $R$ for $\Delta = +20\Gamma$ and $I_p = 126\,$mW/cm$^2$ (black) and $\Delta = +40\Gamma$ and $I_p = 250\,$mW/cm$^2$ (blue). Values above one indicate violation of a Cauchy-Schwarz inequality for classical fields. Dotted lines provide maximum and minimum values of Eq.~\eqref{g12}, while solid lines are fits to Eq.~\eqref{g12b}.}
    \label{Fig3}
\end{figure}

Figure~\ref{Fig3}a, then, plots all correlation functions required to measure $R(\tau)$ for the parameters of Fig.~\ref{Fig2} and up to $|\tau| = 30\,$ns. We observe the same oscillatory behaviour for the functions $g_{12}(\tau)$ as predicted in Eq.~\eqref{g12}. The dashed lines providing the same reference values of Fig.~\ref{Fig2}c, and the experimental values of $\bar{g}_{12}(\tau)$ clearly exceeds the maximum prediction for Eq.~\eqref{g12}. The results for the violation of the Cauchy-Schwarz inequality are shown in Fig.~\ref{Fig3}b for two different pairs of detuning and pump intensity: ($+20\,\Gamma$, 126$\,$mW/cm$^2$) and ($+40\,\Gamma$, 250$\,$mW/cm$^2$). For $\Delta = +20\,\Gamma$, we observed a maximum $R_{max} =  (1.98\pm0.03) \nleq 1$, a violation by 33 standard deviations. The dashed lines again provide some reference values: $R=0.25$ for no cross-correlation, $R=1$ for the maximum classical correlations, and $R=1.71$ for the maximum value predicted by Eq.~\eqref{g12}. Again, violations larger than predicted by Eq.~\eqref{g12} are observed.

For a better comparison with the experimental data, we had to empirically modify Eq.~\eqref{g12} to
\begin{equation}
g_{12}^{e}(\tau) = 1 + \frac{4f}{\pi^2} \left[ 1 + {\rm e}^{-\chi \Gamma |\tau|} - 2 \cos \left( \Delta^{\prime} |\tau|\right) {\rm e}^{-\chi \Gamma |\tau|/2} \right] \,, \label{g12b}
\end{equation}
with $f$, $\chi$, and $\Delta^{\prime}$ as fitting parameters. The solid line in Fig.~\ref{Fig3}a provides the fit to Eq.~\eqref{g12b}, resulting in $\Delta^{\prime} = (21.53 \pm 0.07)\,\Gamma$, $f = 1.58\pm 0.01$, and $\chi = 5.1\pm 0.1$. The value $\Delta^{\prime} \approx \Delta$ is consistent with Eq.~\ref{g12}. The value $f>1$, on the other hand, indicates stronger correlations. The most striking difference, however, is the enhanced natural decay represented by the value of $\chi > 1$, which may indicate the occurrence of collective enhancement in the system, as observed in other experiments under similar conditions~\cite{OrtizGutierrez2018,Araujo2016}. Such collective enhancement may also be related to the observation of $f>1$, since it also affects the directionality of the emitted photons~\cite{deOliveira2014}. Using the experimental values for the auto-correlation function at $\tau = 0$ and Eq.~\eqref{R}, the solid line in Fig.~\ref{Fig3}a results in the red solid line in Fig.~\ref{Fig3}b. A similar procedure is used for the green solid line for the $\Delta = +40\,\Gamma$ data. In this case, we obtain $\Delta^{\prime} = (39.5 \pm 0.1)\,\Gamma$, $f = 1.53\pm 0.03$, and $\chi = 5.6\pm 0.3$.

\vspace{-0.3cm}

\begin{figure}[h]
    \centering
    \includegraphics[scale=0.4]{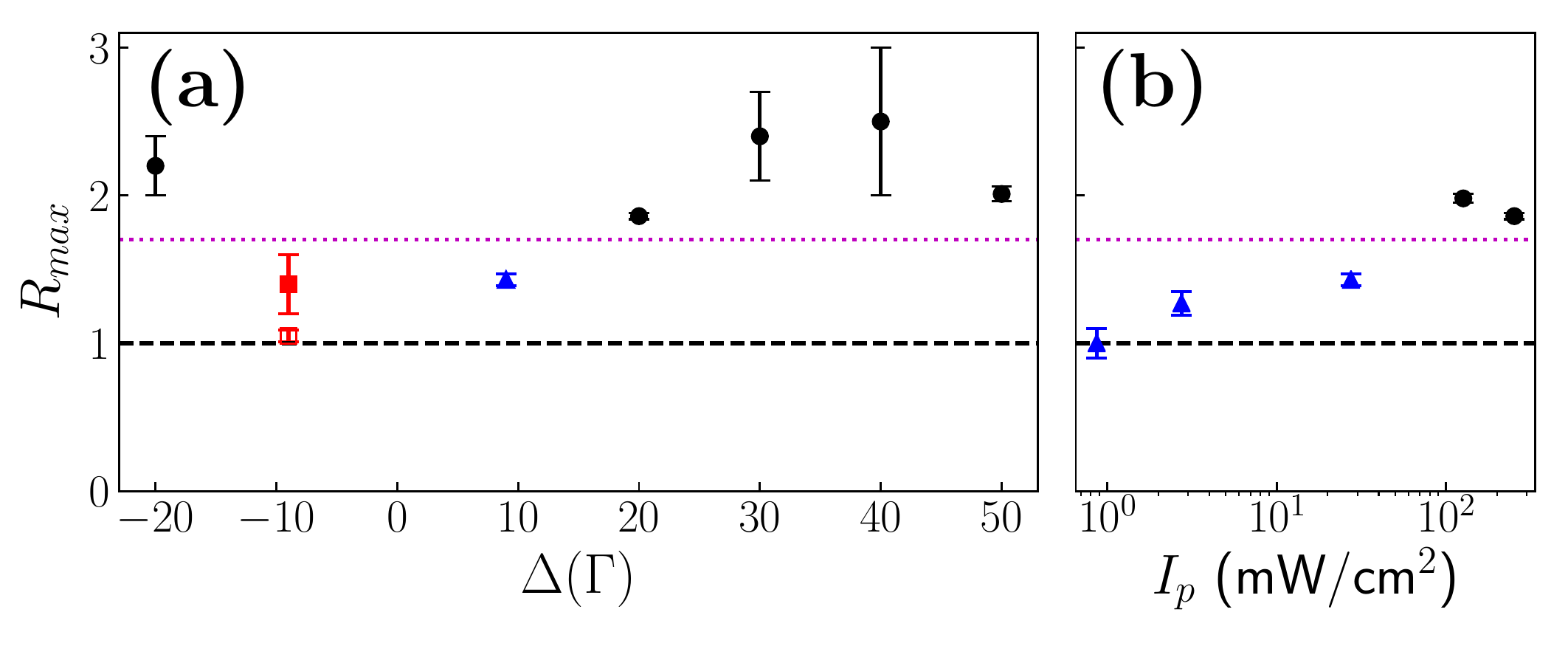}
        \vspace{-0.7cm}
        \caption{$R_{max}$ as a funcion of (a) detuning and (b) pump intensity. For (a), we have $I_p = 250\,$mW/cm$^2$ (black filled circles), $14\,$mW/cm$^2$ (red filled square), and $27\,$mW/cm$^2$ (blue filled triangle), respectively. The empty square is measured for the same parameters as the filled one, but with the MOT repumper laser turned on continuously. For (b), the detunings are $\Delta = +9\,\Gamma$ (blue filled triangles) and $+20\,\Gamma$ (black filled circles). Horizontal lines are the same as for Fig.~\ref{Fig3}b.}
        \label{Fig4}
\end{figure}

$R_{max}$ values measured by this procedure for various parameters are plotted in Fig.~\ref{Fig4}. The filled black circles were obtained by changing the detuning with a fixed pump intensity of $I_p=250\,$mW/cm$^2$. For this experimental sequence, the largest scattering rate is obtained for $\Delta = +20\,\Gamma$, and involved detection rates already close (one tenth) to the saturation rate of our detectors. For smaller detunings, we had then to decrease the intensity of the excitation fields. For $\Delta = +9$ and $-9\,\Gamma$ in Fig.~\ref{Fig4}a, we used $I_p = 27$ and $14\,$mW/cm$^2$, respectively. The $R_{max}$ values for $\Delta = \pm 9\,\Gamma$ are given by the filled triangle and square, respectively, and represent our optimised values for these detunings. The open square gives the result for the same conditions as the filled square, but with the repumper laser turned on continuously. We observe, then, that the repumper laser substantially decreases the observed correlation, leaving it at the border of the classical region. Figure~\ref{Fig4}b plots some data changing the intensity for two different detunings ($\Delta = +9\,\Gamma$ and $+20\,\Gamma$) representing quite different conditions of $R_{max}$. In general, we observe a robust violation $R_{max} \nleq 1$ for a broad variation of parameters without large variations of its value. This overall behaviour is in agreement with Eq~\eqref{g12}, even though the specific shape of the curves and maximum correlations were best approximated by the empirical Eq.~\eqref{g12b}. In practice, we found $R_{max}$ mostly affected by alignment of the four-wave-mixing fields, polarisation optimisation, and elimination of other background fields, like the MOT repumper laser. Degradation on these general conditions would set $R_{max} \leq 1$. However, once the conditions are set for the correlations to rise above the classical limit, they become quite insensitive and robust to large variations of key parameters.

In conclusion, we demonstrated experimentally the nonclassical correlation between photon pairs emitted through SFWM in an ensemble of pure two-level atoms, confirming previous theoretical predictions~\cite{Du2007,Wen2007}. We also demonstrated the broad independence of the correlation to various experimental parameters that commonly control the purity of photon pairs generated from atomic ensembles, like the number of atoms in the sample and the detection rates (determined by detuning and excitation intensity). We were able to observe clear nonclassical correlations up to detection rates close to the saturation of our photodetectors. The experimental data, however, also reveal discrepancies with the theoretical predictions, showing larger nonclassical correlations and faster decays, which are currently under investigation. These observations point to the possibility of observing and exploring nonclassical correlations in more mundane situations, over strong backgrounds of uncorrelated photons, stressing the importance of developing complete quantum treatments for situations commonly believed to be well covered by semi-classical theories. Finally, the reported nonclassical correlation may be considerably enhanced by filtering the component of Rayleigh scattering and exploring multi-partite correlations to other directions. 

\begin{acknowledgments}
This work was supported by the Brazilian funding agencies CNPq (Conselho Nacional de Desenvolvimento Cient\'{\i}fico e Tecnol\'{o}gico), CAPES (Coordena\c{c}\~{a}o de Aperfei\c{c}oamento de Pessoal de N\'{\i}vel Superior), and FACEPE (Funda\c{c}\~ao de Amparo \`{a} Ci\^{e}ncia e Tecnologia do Estado de Pernambuco), through the programs PRONEM and INCT-IQ (Instituto Nacional de Ci\^{e}ncia e Tecnologia de Informa\c{c}\~{a}o Qu\^{a}ntica).

M. O. A. and L. S. M. contributed equally to this work. 
\end{acknowledgments}


\begin{thebibliography}{0}%
\makeatletter
\providecommand \@ifxundefined [1]{%
 \@ifx{#1\undefined}
}%
\providecommand \@ifnum [1]{%
 \ifnum #1\expandafter \@firstoftwo
 \else \expandafter \@secondoftwo
 \fi
}%
\providecommand \@ifx [1]{%
 \ifx #1\expandafter \@firstoftwo
 \else \expandafter \@secondoftwo
 \fi
}%
\providecommand \natexlab [1]{#1}%
\providecommand \enquote  [1]{``#1''}%
\providecommand \bibnamefont  [1]{#1}%
\providecommand \bibfnamefont [1]{#1}%
\providecommand \citenamefont [1]{#1}%
\providecommand \href@noop [0]{\@secondoftwo}%
\providecommand \href [0]{\begingroup \@sanitize@url \@href}%
\providecommand \@href[1]{\@@startlink{#1}\@@href}%
\providecommand \@@href[1]{\endgroup#1\@@endlink}%
\providecommand \@sanitize@url [0]{\catcode `\\12\catcode `\$12\catcode
  `\&12\catcode `\#12\catcode `\^12\catcode `\_12\catcode `\%12\relax}%
\providecommand \@@startlink[1]{}%
\providecommand \@@endlink[0]{}%
\providecommand \url  [0]{\begingroup\@sanitize@url \@url }%
\providecommand \@url [1]{\endgroup\@href {#1}{\urlprefix }}%
\providecommand \urlprefix  [0]{URL }%
\providecommand \Eprint [0]{\href }%
\providecommand \doibase [0]{http://dx.doi.org/}%
\providecommand \selectlanguage [0]{\@gobble}%
\providecommand \bibinfo  [0]{\@secondoftwo}%
\providecommand \bibfield  [0]{\@secondoftwo}%
\providecommand \translation [1]{[#1]}%
\providecommand \BibitemOpen [0]{}%
\providecommand \bibitemStop [0]{}%
\providecommand \bibitemNoStop [0]{.\EOS\space}%
\providecommand \EOS [0]{\spacefactor3000\relax}%
\providecommand \BibitemShut  [1]{\csname bibitem#1\endcsname}%
\let\auto@bib@innerbib\@empty
\end{thebibliography}%


%


\begin{thebibliography}{16}

\bibitem{Yariv1989}
A. Yariv, {\it Quantum Electronics} (John Wiley \& Sons, New York, 1989).

\bibitem{Slusher1985}
R. E. Slusher, L. W. Hollberg, B. Yurke, J. C. Mertz, and J. F. Valley, Observation of Squeezed States Generated by Four-Wave Mixing in an Optical Cavity, Phys. Rev. Lett. {\bf 55}, 2409 (1985).

\bibitem{Maeda1987}
Mari W. Maeda, Prem Kumar, and Jeffrey H. Shapiro, Observation of squeezed noise produced by forward four-wave mixing in sodium vapor, Opt. Lett. {\bf 32}, 178 (1987).

\bibitem{Raizen1987}
M. G. Raizen, L. A. Orozco, Min Xiao, T. L. Boyd, and H. J. Kimble, Squeezed-State Generation by the Normal Modes of a Coupled System, Phys. Rev. Lett. {\bf 59}, 198 (1987).

\bibitem{Vallet1990}
M. Vallet, M. Pinard, and G. Grynberg, Generation of Twin Photon Beams in a Ring Four- Wave Mixing Oscillator, Europhys. Lett. {\bf 11}, 739 (1990).

\bibitem{Lambrecht1996}
A. Lambrecht, T. Coudreau, A. M. Steinberg, and E. Giacobino, Squeezing with cold atoms, Europhys. Lett. {\bf 36}, 93 (1996).

\bibitem{Ries2003}
J. Ries, B. Brezger, and A. I. Lvovsky, Experimental vacuum squeezing in rubidium vapor via self-rotation, Phys. Rev. A {\bf 68}, 025801 (2003).

\bibitem{McCormick2007}
C. F. McCormick, V. Boyer, E. Arimondo, and P. D. Lett, Strong relative intensity squeezing by four-wave mixing in rubidium vapor, Opt. Lett. {\bf 32}, 178 (2007).

\bibitem{Duan2001}
L.-M. Duan, M. D. Lukin, J. I. Cirac, and P. Zoller, Long-distance quantum communication with atomic ensembles and linear optics, Nature {\bf 414}, 413 (2001).

\bibitem{Kuzmich2003}
A. Kuzmich, W. P. Bowen, A. D. Boozer, A. Boca, C. W. Chou, L.-M. Duan, and H. J. Kimble, Generation of non-classical photon pairs for scalable quantum communication with atomic ensembles, Nature 423, 731 (2003).

\bibitem{Balic2005}
V. Balic, D. A. Braje, P. Kolchin, G. Y. Yin, and S. E. Harris, Generation of Paired Photons with Controllable Waveforms, Phys. Rev. Lett. {\bf 94}, 183601 (2005).

\bibitem{Matsukevich2005}
D. N. Matsukevich, T. Chaneli\`ere, M. Bhattacharya, S.-Y. Lan, S. D. Jenkins, T. A. B. Kennedy, and A. Kuzmich, Entanglement of a Photon and a Collective Atomic Excitation, Phys. Rev. Lett. {\bf 95}, 040405 (2005).

\bibitem{Laurat2006}
J. Laurat, H. de Riedmatten, D. Felinto, C.-W. Chou, E. W. Schomburg, and H. J. Kimble, Efficient retrieval of a single excitation stored in an atomic ensemble, Opt. Express {\bf 14}, 6912 (2006).

\bibitem{Thompson2006}
J. K. Thompson, J. Simon, H. Loh, and V. Vuleti\'c, A High-Brightness Source of Narrowband, Identical-Photon Pairs, Science {\bf 313}, 74 (2006).

\bibitem{Zhao2009}
B. Zhao, Y.-A. Chen, X.-H. Bao, T. Strassel, C.-S. Chuu, X.-M. Jin, J. Schmiedmayer, Z.-S. Yuan, S. Chen, and J.-W. Pan, A millisecond quantum memory for scalable quantum networks,  Nature Phys. {\bf 5}, 95 (2009).

\bibitem{Albrecht2015}
B. Albrecht, P. Farrera, G. Heinze, M. Cristiani, and H. de Riedmatten, Controlled Rephasing of Single Collective Spin Excitations in a Cold Atomic Quantum Memory, Phys. Rev. Lett. {\bf 115}, 160501 (2015).

\bibitem{OrtizGutierrez2018}
L. Ortiz-Guti\'errez, L. F. Mu\~noz-Mart\'inez, D. F. Barros, J. E. O. Morales, R. S. N. Moreira, N. D. Alves, A. F. G. Tieco, P. L. Saldanha, and D. Felinto, Experimental Fock-State Superradiance, Phys. Rev. Lett. {\bf 120}, 083603 (2018).

\bibitem{Chou2005}
C. W. Chou, H. de Riedmatten, D. Felinto, S. V. Polyakov, S. J. van Enk, and H. J. Kimble, Measurement-induced entanglement for excitation stored in remote atomic ensembles, Nature {\bf 438}, 828 (2005).

\bibitem{Chou2007}
C. W. Chou, J. Laurat, H. Deng, K. S. Choi, H. de Riedmatten, D. Felinto, and H. J. Kimble, Functional quantum nodes for entanglement distribution over scalable quantum networks, Science {\bf 316}, 1316 (2007).

\bibitem{Strutt1871}
J. W. Strutt, On the Light from the Sky, its Polarization and Colour, The London, Edinburgh, and Dublin Philosophical Magazine and Journal of Science {\bf 41}, 107 (1871).

\bibitem{Loudon1983}
R. Loudon, {\it The Quantum Theory of Light} (Oxford University Press, New York, 1983).

\bibitem{Moreira2021}
R. S. N. Moreira, P. J. Cavalcanti, L. F. Mu\~noz-Mart\'inez, J. E. O. Morales, P. L. Saldanha, J. W. R. Tabosa, and Daniel Felinto, Nonvolatile atomic memory in the spontaneous scattering of light from cold two-level atoms, Opt. Commun. {\bf 495}, 127075 (2021).

\bibitem{Du2007}
S. Du, J. Wen, M. H. Rubin, and G. Y. Yin, Four-Wave Mixing and Biphoton Generation in a Two-Level System, Phys. Rev. Lett. {\bf 98}, 053601 (2007).

\bibitem{Wen2007}
J. Wen, S. Du, and M. H. Rubin, Biphoton generation in a two-level atomic ensemble, Phys. Rev. A {\bf 75}, 033809 (2007).

\bibitem{Wen2008}
J. Wen, S. Du, M. H. Rubin, and E. Oh, Two-photon beating experiment using biphotons generated from a two-level system, Phys. Rev. A {\bf 78}, 033801 (2008).

\bibitem{Felinto2005}
D. Felinto, C. W. Chou, H. de Riedmatten, S. V. Polyakov, and H. J. Kimble, Control of decoherence in the generation of photon pairs from atomic ensembles, Phys. Rev. A {\bf 72}, 053809 (2005).

\bibitem{deOliveira2014}
R. A. de Oliveira, M. S. Mendes, W. S. Martins, P. L. Saldanha, J.W.R. Tabosa, and D. Felinto, Single-photon superradiance in cold atoms, Phys. Rev. A {\bf 90}, 023848 (2014).

\bibitem{Allen1987}
L. Allen and J. H. Eberly, {\it Optical Resonance and Two-Level Atoms} (Dover Publications, New York, 1987).

\bibitem{Lopez2019}
J. P. Lopez, A. M. G. de Melo, D. Felinto, and J. W. R. Tabosa, Observation of giant gain and coupled parametric oscillations between four optical channels in cascaded four-wave mixing, Phys. Rev. A {\bf 100}, 023839 (2019).

\bibitem{Capella2021}
J. C. C. Capella, A. M. G. de Melo, J. P. Lopez, J. W. R. Tabosa, and D. Felinto, Atomic memory based on recoil-induced resonances, arXiv:2112.14800 (2021).

\bibitem{Clauser1974}
J. F. Clauser, Experimental distinction between the quantum and classical field-theoretic predictions for the photoelectric effect, Phys. Rev. D {\bf 9}, 853 (1974).

\bibitem{deAlmeida2016}
A. J. F. de Almeida, M.-A. Maynard, C. Banerjee, D. Felinto, F. Goldfarb J. W. R. Tabosa, Phys. Rev. A {\bf 94}, 063834 (2016).

\bibitem{SM}
See Supplemental Material at SMAraujo2022.pdf for additional details on optical pumping in our system.

\bibitem{Eloy2018}
A. Eloy, Z. Yao, R. Bachelard, W. Guerin, M. Fouch\'e, and R. Kaiser, Diffusing-wave spectroscopy of cold atoms in ballistic motion, Phys. Rev. A {\bf 97}, 013810 (2018).

\bibitem{Araujo2016}
M. O. Ara\'ujo, I. Kre\v{s}i\'c, R. Kaiser, and W. Guerin, Superradiance in a Large and Dilute Cloud of Cold Atoms in the Linear-Optics Regime, Phys. Rev. Lett. {\bf 117}, 073002 (2016).

\end{thebibliography}
\end{document}


\title{Supplemental Material for `Observation of nonclassical correlations in biphotons \\ generated from an ensemble of pure two-level atoms'}

\author{Michelle O. Ara\'ujo, Lucas S. Marinho, and Daniel Felinto}
 \affiliation{Departamento de F\'{\i}sica, Universidade Federal de Pernambuco, 50670-901, Recife, Pernambuco, Brazil}

\date{\today}

\begin{abstract}
Here we provide additional details about the optical pumping process to a single Zeeman sub-level.
\end{abstract}

\pacs{Valid PACS appear here}

\maketitle

The photon-pair generation by spontaneous four-wave mixing (SFWM) investigated in this work is a parametric process, which implies that the material system ends the process at the same level it started. This is crucial for the collective constructive interference behind the dependence with $N^2$ of the nonlinear process, with $N$ the number of atoms participating in the process. In our experiment, we prepare the atoms initially in the hyperfine state $5S_{1/2}(F=2)$ with five Zeeman sub-levels. If the atomic populations were equally distributed among these sub-levels, our signal would be proportional to $\sum_{m_F} (N/5)^2 = N^2/5$, instead of the $N^2$ if all atoms were at the same Zeeman sub-level~\cite{Felinto2005}. In order for the SFWM to prevail over the background of Rayleigh scattering, it is crucial, then, to optically pump all atoms to a single Zeeman sub-level. We do that by using pumping beams with $\sigma^+$ polarisations, which optically pump the atoms to the $m_F = +2$ sub-level. 

After being pumped to $m_F=+2$, the atoms can only be excited at the $\sigma^+$ transition leading to $5P_{3/2}(F^{\prime}=3,m_{F^{\prime}}=+3)$ and scatter light only with $\sigma^+$ polarisation. A straightforward way to check the quality of the optical pumping (and to adjust the polarisation of the pumping beams) is then to measure the polarisation of the scattered light. As reported in the main text, we measure ($99.0\pm 0.2$)\% for the scattered light's degree of polarisation. This degree of polarisation, however, is not the most sensitive measurement of the degree of polarisation of the atomic ensemble itself, since atoms at $m_F = +1$ cannot generate $\sigma^-$ light as well.

\renewcommand{\thefigure}{S1}
\begin{figure}[th]
    \centering
   \includegraphics[scale=0.30]{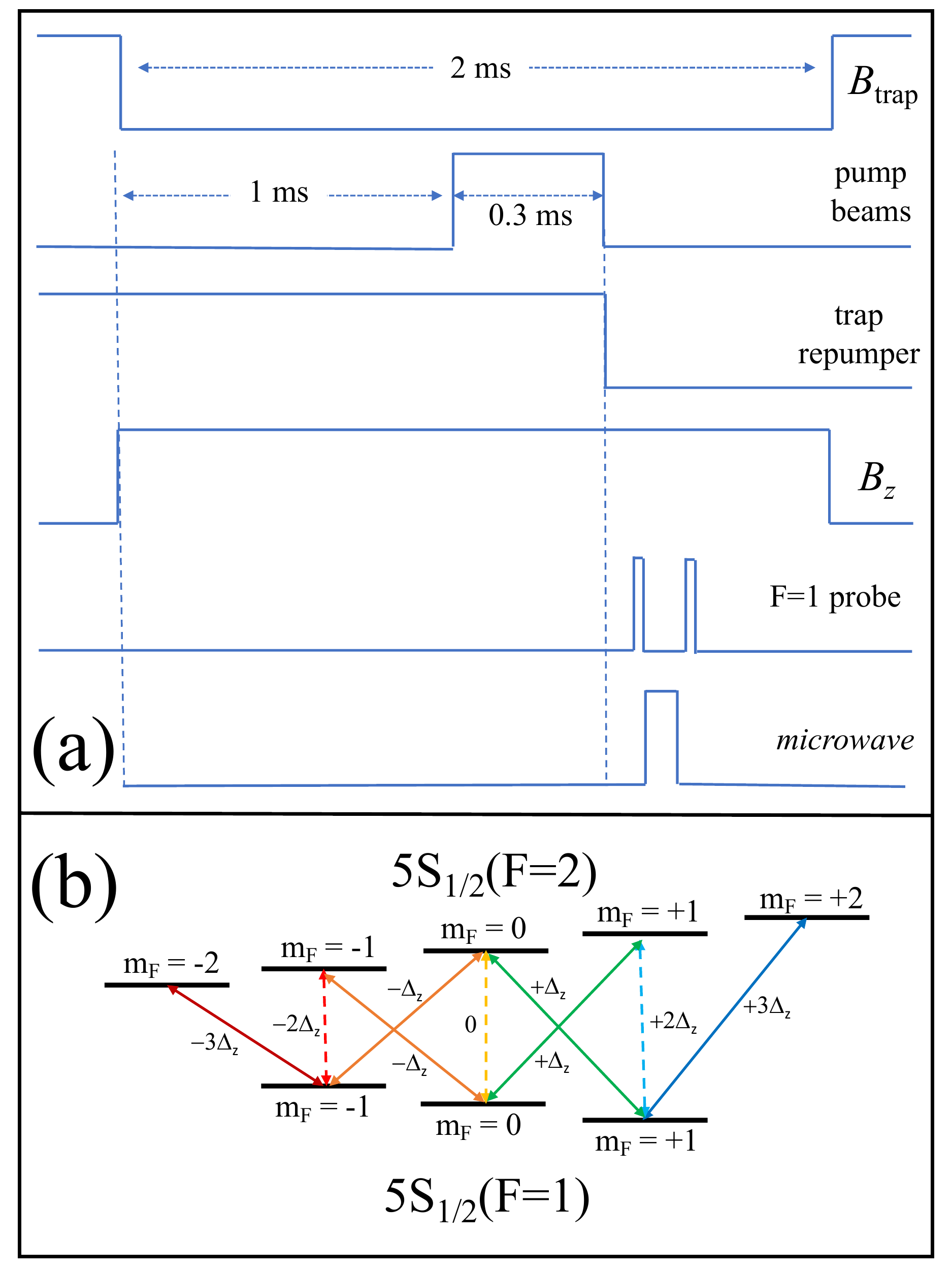}
        \caption{(a) Time sequence for magnetic and laser fields involved in the measurement of the microwave spectrum. (b) Possible transitions between Zeeman sub-levels of the two hyperfine ground states. The transitions favoured by our antenna configuration are indicated by solid arrows.}
    \label{Fig1}
\end{figure}

In order to directly address the distribution of atoms at the different sub-levels resulting from optical pumping, we performed a microwave spectroscopy of the Zeeman structure of the atomic ensemble~\cite{deAlmeida2016} with and without the optical pumping. This microwave spectroscopy is also the technique we use to cancel, using three sets of bias coils, the spurious magnetic field that is left after turning off the trap's magnetic field ($B_{trap}$). In Fig.~\ref{Fig1}a we show the time sequence of our procedure to perform microwave spectroscopy at various situations. The simplest one is to measure the distribution of Zeeman sub-levels resulting from the spurious magnetic field in the trap's region. In this case, we turn off $B_{trap}$ for 2~ms, but keep the repumper light on for a while to prepare the atoms at the $5S_{1/2}(F=2)$ state. After turning off the repumper, we probe the atoms with a 1-$\mu$s long pulse resonant with the $5S_{1/2}(F=1) \rightarrow 5P_{3/2}(F^{\prime}=2)$ transition. There is no atoms at the $5S_{1/2}(F=1)$ state and this first pulse is used just to calibrate the detection of a second probe pulse for power fluctuations of the probe laser. After this calibration pulse, we finally excite the sample with a 25-$\mu$s long microwave pulse of frequency (6.834682 + $f_{mw}$) GHz, with $f_{mw}$ the frequency difference to the resonance of the magnetically insensitive transition $5S_{1/2}(F=1,m_F=0) \rightarrow 5S_{1/2}(F=2,m_F=0)$. The microwave, then, transfers population to the $F=1$ state if $f_{mw}$ reaches a resonance with a populated Zeeman sub-level of $F=2$. After the microwave pulse, a second probe pulse reaches the ensemble and its degree of absorption (optical depth $OD_{mw}$) indicates the level of population at the original $F=2$ Zeeman sub-level. Our microwave spectrum is then a plot of $OD_{mw}$ versus $f_{mw}$. In Fig.~\ref{Fig2}a, we have the microwave spectrum resulting from the cancellation of the spurious magnetic field in our sample's region. It indicates a spurious magnetic field of approximately 20~mG, similar to the one reported in Ref.~\onlinecite{Moreira2021}.

After cancelling the spurious field as much as possible in our setup, we then turn on a magnetic field $B_z$ in the direction of the pump beam, our $z$ quantisation axis, during the period in which $B_{trap} = 0$. This $B_z$ field, then, splits the spectrum in various peaks (Fig.~\ref{Fig2}b) corresponding to transitions between different Zeeman sub-levels, as shown in Fig.~\ref{Fig1}b. The microwave pulse is generated by an antenna positioned outside the vacuum chamber and aligned approximately with the $z$ axis. In this way, the microwave favours $\sigma^+$ and $\sigma^-$ transitions between the Zeeman sub-levels of the ground hyperfine states. In Fig.~\ref{Fig1}b, the favoured transitions are indicated by the arrows with solid lines, while the suppressed $\pi$ transitions are indicated by arrows with dashed lines. Since the Land\'e $g_F$-factors for the two hyperfine states have approximately the same amplitudes but opposite signs, two of the transitions are degenerate (indicated by the orange and green arrows). If $|f_{mw}| = \Delta_z$ gives the position of the first peak outside the origin of the spectrum, our antenna position favours then the peaks with $f_{mw} = \pm \Delta_z\,,\pm 3\Delta_z$, as highlighted in Fig.~\ref{Fig2}b.

\renewcommand{\thefigure}{S2}
\begin{figure}[th]
    \centering
   \includegraphics[scale=0.40]{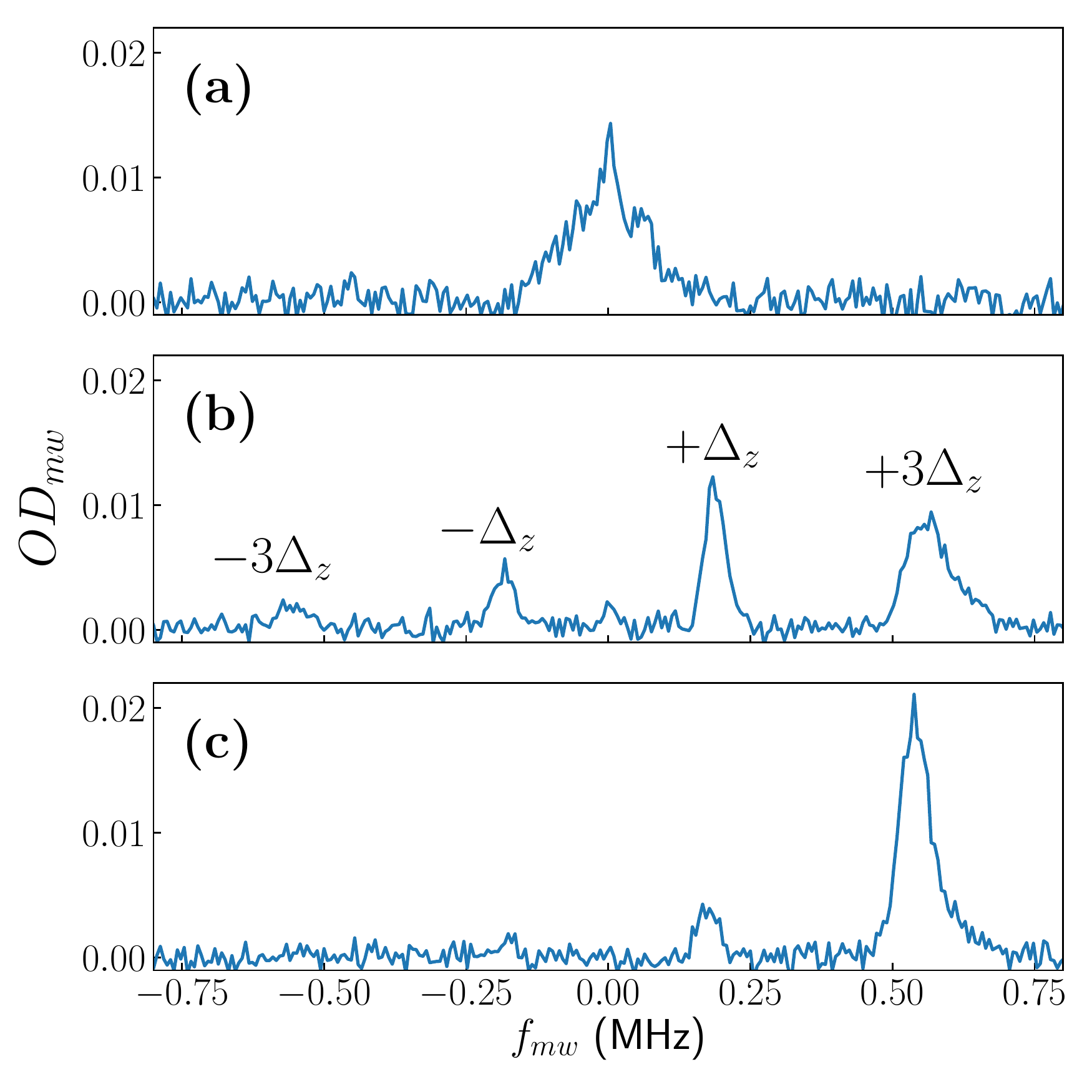}
   \vspace{-0.4cm}
        \caption{(a) Microwave spectrum of the residual magnetic field in the sample's region, after procedure to optimise its cancelation. (b) Microwave spectrum of panel (a) split in its Zeeman components by a continuous magnetic field applied in the $z$ direction. (c) Modified spectrum after optical pumping to $m_F = +2$.}
    \label{Fig2}
\end{figure}

In Fig.~\ref{Fig2}c, we finally present the microwave spectrum after the optical pumping beams are turned on for 300$\,\mu$s, as shown in Fig.~\ref{Fig1}a. In this was, we can see directly the accumulation of population in the $m_F = +2$ sub-level, with some residual population still at $m_F=+1$. This residual population at $m_F=+1$ is the origin of the spurious optical pumping in Figs.~2a and~2b of the main text, as explained there, since it may couple to the  $5P_{3/2}(F^{\prime}=2,m_F^{\prime}=+2)$ and, from there, decay to the $F=1$ dark ground state. This way of characterising the optical pumping to the $m_F = +2$ sub-level is more sensitive than looking at the scattered light polarisation, in the sense that decreasing the $+\Delta_z$ peak allows for a finer adjustment of the waveplates controlling the polarisation of the pumping beams. However, this finer optimisation of the optical pumping does not translate in significant increases in the degree of the observed non-classical correlations, with respect to the simpler procedure based in optimisation of the scattered light's polarisation.

The spectrum in Fig.~\ref{Fig2}c allows us to directly check and optimise the atomic polarisation, but it does not provide a sensible quantification of it. The problem is that the distribution of peak heights in the spectra in Fig.~\ref{Fig2} depends not only on the original populations of each Zeeman sub-level, but also on the fine positioning and alignment of the antenna generating the microwave. We can then observe improvements of the optical pumping, but not extract the exact population of each sub-level. Note also in Fig.~\ref{Fig1}a that the measurement of Fig.~\ref{Fig2}c could not be performed at the same conditions of the experiments reported in the main text, since the repumper laser was kept on during the optical pumping to $m_F = +2$. The problem is that the spurious optical pumping to $F=1$, due to the pump beams, affects directly our measurement of the microwave spectrum, since it relies in a probe measuring the overall population in $F=1$. Even though, the above procedure allows for the finest tuning of the polarisation of the pumping beams used in our final experimental configuration.